\documentstyle[titlepage,12pt]{article}
\begin{document}
\renewcommand{\theequation}{\thesection.\arabic{equation}}
\title{Quantum Tunneling Effect in Oscillating Friedmann Cosmology}
\author{Mariusz P. D\c{a}browski\thanks{Institute of Physics, University of
Szczecin, Wielkopolska 15, 70-451 Szczecin, \hspace*{6mm}Poland.} and
Arne L. Larsen\thanks{Observatoire de Paris,
DEMIRM. Laboratoire Associ\'{e} au CNRS
UA 336, Ob- \hspace*{6mm}servatoire de Paris et
\'{E}cole Normale Sup\'{e}rieure. 61, Avenue
de l'Observatoire, \hspace*{6mm}75014 Paris, France.}}
\maketitle
\begin{abstract}
It is shown that the tunneling effect in quantum cosmology is possible not only
at the very beginning or the very end of the evolution, but also at the moment
of maximum expansion of the universe. A positive curvature expanding
Friedmann universe
changes its state of evolution spontaneously and completely,
{\it without} any changes
in the matter content, avoiding recollapse, and falling into
oscillations between the nonzero values of the scale factor. On the other hand,
an oscillating nonsingular
universe can tunnel spontaneously to a recollapsing regime. The
probability of such kind of tunneling is given explicitly.
It is inversely related to the amount of
nonrelativistic matter (dust), and grows from a certain fixed
value to unity if the negative cosmological constant approaches zero.
\vskip 24pt
\begin{centerline}
{\bf PACS number(s): 04.20.Jb; 04.60.+n; 98.80.Cq; 98.80.Hw}
\end{centerline}
\end{abstract}

\section{Introduction}
\setcounter{equation}{0}
The idea of unification of all forces in nature is quite old. It seems that
GUT's or, in particular, the strong and electroweak theories
have been accepted as standard
theories of physics. However, the most interesting problem today is the
unification of quantum mechanics with Einstein's theory of
gravity. Some fundamentals of
such a unification were first developed in the 60's by DeWitt, Wheeler and
Misner \cite{dew,whe,mis}.
The idea that the universe could have been created through a quantum
mechanical tunneling process from the vacuum, appeared in the 70's
\cite{alb,try,bro}.
More interest was concentrated on the topic in the
80's, mainly because of the emergence of the idea of inflation \cite{gu}.
Especially, Atkatz and Pagels \cite{atp} considered
a possibility for the universe to tunnel quantum mechanically from
an initial static state, while Vilenkin \cite{vil} considered the
tunneling from literally "nothing". Almost simultaneously,
Hartle and Hawking \cite{hh}
developed a
formalism of quantum gravity which is based on the Feynman path integral
approach.

Recently, D\c{a}browski \cite{dab} considered a class of Friedmann cosmologies
which are oscillating, i.e., in which the universe oscillates between the
nonzero values of the scale factor for
infinitely long time. These solutions, in a
way, generalize the static Einstein solution since they require
some balance between gravitational attraction and repulsion. In his
original paper \cite{ein}, Einstein
admitted nonrelativistic matter as a source of attraction and a positive
cosmological constant as a source of repulsion. In our approach the problem is
more complex since we admit more sources corresponding to
either attraction or repulsion. These
are: nonrelativistic matter (attraction), the negative cosmological constant
(attraction) and domain-wall-like matter (repulsion). In general, we can
admit also radiation pressure (attraction), the positive cosmological
constant (repulsion) and string-like-matter (repulsion), but they do not
lead to any qualitative changes for the oscillating solutions \cite{dab}.
It should be stressed that the presence of domain-wall-like matter is essential
for the existence of the nonsingular oscillating universes
considered in this paper.
Being quite a good candidate for the dark matter, stable domain walls are
often considered to be cosmologically
"disastrous" \cite{zel}. However, Hill {\it et} {\it al} \cite{hill} introduced
the socalled "light" or "soft" (and thick)
domain walls in a post-decoupling phase
transition scheme, and found that the produced domain walls were not
necessarily in contradiction with the observed large scale structure of
the universe (see also Ref.\cite{gel,go}). We also point out that strings
and walls could have played a more important role in the early universe, where
the tunneling processess under consideration here could be relevant, than they
do today. Notice also that there is a way to avoid the
domain walls by replacing them by a kinematically equivalent scalar field
\cite{dab}, which is postulated to be another source of the dark matter in
the present era of the universe \cite{mcd,ba}.
For more discusion on the string-like and wall-like matter, see
Ref.\cite{dab}

Usually, when dealing with quantum tunneling of the universe, all
the quantum effects are considered
to appear either at the very beginning or at the very end
of the evolution. Recently, Kiefer and Zeh \cite{kieze} supported an earlier
idea of Gold \cite{gold}, that quantum effects for the universe as a whole may
also be very important at the moment of its maximum expansion (i.e., when the
universe changes its dynamics from expansion to collapse). Our work follows
the spirit of these references and is most naturally related to the early paper
by Atkats and Pagels \cite{atp}, in the sense that we do not consider the
quantum creation of the universe as a whole out of "nothing", but the quantum
tunneling of the universe from one state to another. An important result of
our analysis is that the evolution of the universe can change character
completely {\it without} any changes in the matter content. This should be
contrasted with ordinary cosmological models where dramatic continuous
(or discontinuous) changes in the scale factor usually follow from dramatic
continuous (or discontinuous) changes
in the matter content. Notice, in particular,
the paper by Hawking and Moss \cite{mos}, where a tunneling of the
scalar field, equivalent to a discontinuous change in the matter content,
leads to a spontaneous discontinous change of the Hubble parameter, i.e., of
the evolution of the universe.

The plan of the paper is as follows. In Section 2 we comment on the classically
oscillating Friedmann cosmologies discussed previously in Ref.\cite{dab}. Then
we give a particular solution in terms of Jacobian Elliptic Functions,
useful for the
further purposes of this work. In Section 3 we use WKB approximation in order
to
calculate the tunneling probability of the Friedmann universe from the
expanding
regime to the oscillating regime and from the oscillating regime to the
recollapsing regime, for the particular solution given in Section 2. In Section
4 we discuss the results.

\section{Classically Oscillating Friedmann Cosmologies}
\setcounter{equation}{0}
The existence of oscillating nonsingular solutions of the Friedmann equation
was first mentioned by Harrison \cite{har}. He investigated qualitatively the
universes filled with different kinds of both negative and positive pressure
matter following the early discussion by Robertson \cite{rob} in the 30's.
Recently, oscillating Friedmann cosmologies
have been discussed qualitatively in the
context of "exotic" matter by Kardashev \cite{kar}. A complete discussion
of the oscillating solutions in terms of the Weierstrass Elliptic Functions
has been given by D\c{a}browski \cite{dab}. In this section we will
consider in more detail a family of the simplest, but still
nontrivial, solutions found in
there. To fix our notation, and for later
use, we first review a few general aspects of the construction.

The nonvanishing components of the Ricci tensor for the generic
four-dimensional Friedmann-Robertson-Walker line element, in comoving
coordinates,
\begin{equation}
ds^2=-dt^2+a^2(t)\frac{d\vec{x}d\vec{x}}{(1+\frac{K}{4}\vec{x}\vec{x})^2},
\end{equation}
are given by
\begin{equation}
R_{tt}=-3\frac{a_{,tt}}{a},\;\;\;\;\;\;\;R_{ii}=\frac{2K+aa_{,tt}+
2a^2_{,t}}{(1+\frac{K}{4}\vec{x}\vec{x})^2},
\end{equation}
where $a=a(t)$ is the scale factor, $a_{,t}\equiv da/dt,\;\;
a_{,tt}\equiv d^2a/dt^2\;$ and $K$ is the curvature index. The
corresponding scalar curvature is
\begin{equation}
R=\frac{6}{a^2}(aa_{,tt}+a^2_{,t}+K).
\end{equation}
Including a nonvanishing cosmological constant $\Lambda,$ the action takes
the form
\begin{equation}
S=S_{g}+S_{m}=\frac{1}{16\pi G}\int\sqrt{-g}\;(R-2\Lambda)d\Omega+S_{m},
\end{equation}
and the matter energy-momentum tensor $T_{\mu\nu}$ is defined through
\begin{equation}
\delta S_{m}=-\frac{1}{2}\int\sqrt{-g}\;T_{\mu\nu}\delta g^{\mu\nu}d\Omega.
\end{equation}
In Section 3, we shall return to the explicit expression for $S_{m};$ at this
point we need only eq.(2.5), which together with the variation of
eq.(2.4), leads to the Einstein
equations in standard form. In the presence of radiation, nonrelativistic
matter, string-like and wall-like matter, with corresponding
densities $C_{r},$ $C_{m},$
$C_{s}$ and $C_{w},$ respectively, the $(tt)$-component of the Einstein
equations becomes \cite{dab,aj}
\begin{equation}
a^2 a^2_{,t}=C_{r}+aC_{m}-a^2(K-C_{s})+a^3C_{w}+\frac{\Lambda}{3}a^4,
\end{equation}
and the $(ii)$-component follows by differentiation with respect to $t.$
Introducing the conformal time, $d\tau\equiv dt/a(t),$ eq.(2.6) can be
written as
\begin{equation}
\left(\frac{dT}{d\tau}\right)^2=\alpha T^4+\frac{2}{3}T^3-(K-C_{s})T^2+\beta T+
\frac{\lambda}{3},
\end{equation}
where
\begin{equation}
\Lambda_c^{-1/2}\equiv\frac{3}{2}C_{m},\;\;\;\;\;
\beta\equiv C_{w}\Lambda_c^{-1/2},\;\;\;\;\;\alpha\equiv C_{r}\Lambda_{c},
\;\;\;\;\;\lambda\equiv\Lambda/\Lambda_{c},
\end{equation}
and the so-called reduced temperature $T$ is defined by
\begin{equation}
T\equiv\frac{\Lambda_c^{-1/2}}{a},
\end{equation}
in agreement with Ref.\cite{dab}. The Einstein equation is now of the
elliptic form and can be explicitly solved in the general case.
As shown in Ref.\cite{dab}, oscillating nonsingular solutions to eq.(2.7)
can be obtained for a variety of parameter values. In the following we will
consider the special, but nontrivial case
\begin{equation}
\alpha=0,\;\;\;K=1,\;\;\;C_{s}=0,\;\;\;\beta=3/8,
\end{equation}
where the
mathematics simplifies
considerably; our results can however be easily extended to the more
general case. We are thus considering a spatially closed universe with
"ordinary" nonrelativistic matter
and "exotic" wall-like matter. Equation (2.7) reduces to
\begin{equation}
\left(\frac{dT}{d\tau}\right)^2=
\frac{2}{3}T^3-T^2+\frac{3}{8}T+\frac{\lambda}{3},
\end{equation}
with the general solution
\begin{equation}
a(\tau)=\frac{\Lambda_{c}^{-1/2}}{6\wp(\tau-\tau_0)+1/2}.
\end{equation}
Here $\wp$ is the Weierstrass Elliptic Function \cite{as} with invariants
\begin{equation}
g_2=\frac{1}{48},\;\;\;\;\;\;\;\;g_3=-\frac{1}{1728}-\frac{\lambda}{108},
\end{equation}
and the discriminant reads as
\begin{equation}
\Delta\equiv g_2^3-27g_3^2=\frac{-\lambda}{432}(\lambda+\frac{1}{8}).
\end{equation}
In the special case $\lambda = - 1/8,$ eq.(2.11) factorizes to \cite{dab}
\begin{equation}
\left(\frac{dT}{d\tau}\right)^2 = \frac{2}{3}\left(T - \frac{1}{4}\right)^2
\left(T - 1\right),
\end{equation}
and $\Delta = 0.$ In this case the solution (2.12) reduces to an expression in
elementary functions \cite{as}.
Oscillating nonsingular solutions are obtained for $\lambda\in\;]-1/8,\;0[\;,$
in which case $\Delta>0.$ The three roots are given by
\begin{equation}
e_1=\frac{1+Z^2}{24\;Z},\;\;e_2=-\frac{(1-i\sqrt{3})}{48\;Z}-
\frac{(1+i\sqrt{3})Z}{48},\;\;
e_3=-\frac{(1+i\sqrt{3})}{48\;Z}-\frac{(1-i\sqrt{3})Z}{48},
\end{equation}
where
\begin{equation}
Z=Z(\lambda)=[-1-16\lambda+4\sqrt{2}\sqrt{\lambda(1+8\lambda)}\;]^{1/3}.
\end{equation}
In the interesting region, $\lambda\in\;]-1/8,\;0[\;,$ they fulfill
\begin{equation}
e_1\geq e_2\geq e_3.
\end{equation}
It is now straightforward to write the solution (2.12) in terms of Jacobian
Elliptic Functions \cite{as,tri}
\begin{equation}
a(\tau)=\frac{\Lambda_{c}^{-1/2}\;
\mbox{sn}^2[\sqrt{e_1-e_3}(\tau-\tau_0)\;|\;(e_2-e_3)/(e_1-e_3)]}{6(e_1-e_3)+
(6e_3+1/2)\mbox{sn}^2[\sqrt{e_1-e_3}(\tau-\tau_0)\;|\;(e_2-e_3)/(e_1-e_3)]}
\end{equation}
The integration constant $\tau_0$ must be carefully chosen to give real
solutions for real $\tau.$ Up to real translations, there are essentially two
possibilities. For $\tau_0=0$ we find the solution
\begin{equation}
a_{(-)}(\tau)=\frac{\Lambda_{c}^{-1/2}\;
\mbox{sn}^2[\sqrt{e_1-e_3}\;\tau\;|\;(e_2-e_3)/(e_1-e_3)]}{6(e_1-e_3)+
(6e_3+1/2)\mbox{sn}^2[\sqrt{e_1-e_3}\;\tau\;|\;(e_2-e_3)/(e_1-e_3)]},
\end{equation}
which oscillates between $a=0$ and $a=a_1,$ where
\begin{equation}
a_1\equiv\frac{\Lambda_{c}^{-1/2}}{6e_1+1/2}.
\end{equation}
Taking instead $\tau_0=iK'[(e_2-e_3)/(e_1-e_3)]/\sqrt{e_1-e_3},\;$ where $K'$
is
the complete elliptic integral of the first kind \cite{as}, we
get the solution
\begin{equation}
a_{(+)}(\tau)=\frac{\Lambda_{c}^{-1/2}}{(6e_3+1/2)+
6(e_2-e_3)\mbox{sn}^2[\sqrt{e_1-e_3}\;\tau\;|\;(e_2-e_3)/(e_1-e_3)]},
\end{equation}
which oscillates between $a=a_2$ and $a=a_3,$ where
\begin{equation}
a_2\equiv\frac{\Lambda_{c}^{-1/2}}{6e_2+1/2},\;\;\;\;\;\;\;\;
a_3\equiv\frac{\Lambda_{c}^{-1/2}}{6e_3+1/2}.
\end{equation}
The physical interpretation of these two solutions,
$a_{(-)}(\tau),\;a_{(+)}(\tau),$ follows by returning to the
original equation of motion (2.6). For the parameter values (2.10), it takes
the form
\begin{equation}
\left(\frac{da}{dt}\right)^2+V(a)=0;\;\;\;\;\;
V(a)=\frac{-2}{3\sqrt{\Lambda_{c}}\;a}+1-
\frac{3\sqrt{\Lambda_{c}}\;a}{8}-\frac{\Lambda_{c}\lambda a^2}{3}.
\end{equation}
Defining the potential in this way, we obtain that the dynamics takes place at
the $a$-axis in a $(a,V(a))$-diagram. For $\lambda\in\;]-1/8,\;0[\;,$ the
potential is shown in Fig.1, which explains the two solutions we found. In this
picture, the universe is inflating ($da/dt>0\;\;\mbox{and}\;\;d^2a/dt^2>0$)
when it is expanding and $dV/da<0.$ The
solution $a_{(-)}(\tau)$ thus describes a deflating universe with scale
factor expanding from $a=0$ to the maximal value $a=a_1,$ and thereafter
re-collapsing to $a=0.$ The solution $a_{(+)}(\tau),$ on the other hand,
describes a nonsingular universe, with scale factor oscillating between
$a=a_2$ and $a=a_3.$ Notice also the behaviour in the two limits
\begin{eqnarray}
&\lambda\rightarrow 0&:\;\;\;\;\;a_1\rightarrow a_2,\;\;a_3\rightarrow\infty,
\nonumber\\
&\lambda\rightarrow -1/8&:\;\;\;\;\;a_2\rightarrow a_3.
\end{eqnarray}
The comoving time of the solutions are obtained from
\begin{equation}
t=\int_0^\tau\;a(\tau')d\tau',
\end{equation}
which leads to
\begin{equation}
t_{(-)}(\tau)=
\frac{\Lambda_{c}^{-1/2}}{6e_3+1/2}\left\{\tau-\frac{1}{\sqrt{e_1-e_3}}
\Pi[\frac{6e_3+1/2}{6(e_3-e_1)};\;\sqrt{e_1-e_3}\;\tau\;|\;
\frac{e_2-e_3}{e_1-e_3}]
\right\},
\end{equation}
\begin{equation}
t_{(+)}(\tau)=\frac{\Lambda_{c}^{-1/2}}{(6e_3+1/2)\sqrt{e_1-e_3}}
\Pi[\frac{6(e_3-e_2)}{6e_3+1/2};\;\sqrt{e_1-e_3}\;\tau\;|\;
\frac{e_2-e_3}{e_1-e_3}],
\end{equation}
where $\Pi$ is the elliptic integral of the third kind, and we are using the
notation of Abramowitz and Stegun \cite{as}. For both solutions, the period
in conformal time is given by
\begin{equation}
P_{\tau}=\frac{2}{\sqrt{e_1-e_3}}K[\frac{e_2-e_3}{e_1-e_3}].
\end{equation}
It is however more relevant to consider the periods in comoving time $t$
\begin{equation}
P_{(-)}=\frac{2\Lambda_{c}^{-1/2}}{(6e_3+1/2)\sqrt{e_1-e_3}}\left\{
K[\frac{e_2-e_3}{e_1-e_3}]-
\Pi[\frac{6e_3+1/2}{6(e_3-e_1)}\;|\;\frac{e_2-e_3}{e_1-e_3}]\right\},
\end{equation}
\begin{equation}
P_{(+)}=\frac{2\Lambda_{c}^{-1/2}}{(6e_3+1/2)\sqrt{e_1-e_3}}
\Pi[\frac{6(e_3-e_2)}{6e_3+1/2}\;|\;\frac{e_2-e_3}{e_1-e_3}].
\end{equation}
In the two extreme limits (2.25) we find
\begin{eqnarray}
\hspace*{-1cm}&\lambda\rightarrow 0&:\;\;P_{(-)}\rightarrow\infty,\;
\;P_{(+)}\rightarrow\infty,
\nonumber\\
\hspace*{-1cm}&\lambda\rightarrow -1/8&:
\;\;P_{(-)}\rightarrow 4\pi\sqrt{2}(2-\sqrt{3})
\Lambda_{c}^{-1/2},\;\;P_{(+)}\rightarrow 8\pi\sqrt{2}\Lambda_{c}^{-1/2}
\end{eqnarray}
It is interesting that the period $P_{(+)}$ is {\it not}
continously going to zero for $\lambda\rightarrow -1/8,$ as could have been
expected from the potential in that particular limit (Fig.1), where the
$a_{(+)}$ solution approaches the Einstein static Universe.
This concludes our discussion of the classical behaviour of the oscillating
solutions, eqs.(2.20),(2.22).
\section{The Quantum Tunneling Probability}
\setcounter{equation}{0}
As it is suggested
from the potential given in Fig.1, although classically stable, the
oscillating nonsingular solution $a_{(+)}$ has the possibility to
tunnel quantum mechanically through the barrier from $a_2$ to $a_1$
and then to collapse into $a=0.$
Similarly, the expanding solution $a_{(-)}$ which classically
recollapses after hitting the barrier,
has the possibility to tunnel quantum mechanically through the barrier from
$a_1$ to $a_2$ and become oscillating and nonsingular. We believe that such
processess could be relevant at least for the early evolution of the universe,
i.e., shortly after the big bang. Notice that we are not discussing here the
quantum creation of the
universe as a whole out of "nothing" \cite{vil}. Our aim is to consider
a spontaneous change of character of
evolution due to quantum processess {\it after} the
big bang. In the standard cosmological models, the evolution of the universe
changes because the matter content changes, essentially
because the temperature changes. What we want to illustrate with our
model in this section is that
there may be an additional effect: without any changes in the matter content,
the evolution of the universe can change completely
and spontaneously due to a quantum mechanical tunneling process.
To be more specific, we will now calculate the probability for the universe to
tunnel through the barrier between $a=a_1$ and $a=a_2,$ Fig.1. In the standard
quantum mechanics calculation, the probability is given in the
WKB-approximation
by (see for instance the recent review by Atkatz, \cite{katz}, with
applications to cosmology)
\begin{equation}
p\sim e^{-B},
\end{equation}
where $B$ is twice the conjugate momentum integrated under the barrier
\begin{equation}
B=2\int_{a_1}^{a_2}\;|P_{a}|da.
\end{equation}
Denoting by $L,$ $L_{g}$ and $L_{m},$ respectively, the Lagrangians
corresponding to $S,$ $S_g$ and $S_{m},$ and assuming that the matter
Lagrangian depends on the scale factor only via $a,$ and not its derivatives
(which is usually the case), we find
\begin{equation}
P_{a}\equiv
\frac{\partial L}{\partial a_{,t}}=\frac{\partial L_{g}}{\partial a_{,t}},
\end{equation}
thus we do not need an explicit expression for the matter Lagrangian to
calculate the tunneling probability (as a check, we calculate
in Appendix A the
tunneling probability using the instanton method where $S_{m}$
is needed explicitly).
The gravitational Lagrangian is
obtained from eqs.(2.3),(2.4), for $K=1$
\begin{equation}
S_{g}=\frac{3\pi}{4G}\int dt[-aa^2_{,t}+a-\Lambda a^3/3],
\end{equation}
so that
\begin{equation}
P_{a}=-\frac{3\pi}{2G}aa_{,t}.
\end{equation}
For the parameter values (2.10), the Einstein equation (2.6) now takes the form
\begin{equation}
P^2_{a}=\frac{1}{\Lambda_{c}}(\frac{3\pi}{2G})^2
[\frac{2}{3}(\sqrt{\Lambda_{c}}\;a)-(\sqrt{\Lambda_{c}}\;a)^2+
\frac{3}{8}(\sqrt{\Lambda_{c}}\;a)^3+
\frac{\lambda}{3}(\sqrt{\Lambda_{c}}\;a)^4].
\end{equation}
Then eq.(3.2) becomes
\begin{equation}
B=\frac{3\pi}{\Lambda_{c}G}\sqrt{\frac{-\lambda}{3}}\int_{x_1}^{x_2}
\sqrt{x(x-x_1)(x_2-x)(x_3-x)}\;dx,
\end{equation}
where $x_{i}\equiv\sqrt{\Lambda_{c}}\;a_{i}\;;\;\;i=1,2,3.$ The integral is of
elliptic type and the result can be expressed in terms of complete elliptic
integrals \cite{as,byrd}, see Appendix B.
For $\lambda\rightarrow 0,$ where $a_1\rightarrow a_2,$ we find as expected
$B=0.$ There is no barrier in that limit so the "tunneling probability"
is unity, but it actually takes
infinite comoving
time for the universe to reach the point $a=a_1=a_2$ (from below or
above). In the other extreme limit, $\lambda\rightarrow -1/8,$ where
the solution $a_{(+)}$ is static, the integral (3.7) factorizes to give
\begin{equation}
B(\lambda=-1/8)=\frac{\pi}{2G}\sqrt{\frac{3}{2}}
\int_{\Lambda_{c}^{-\frac{1}{2}}}^{4\Lambda_{c}^{-\frac{1}{2}}}
(4-a\Lambda_{c}^{\frac{1}{2}})
\sqrt{a( a-\Lambda_{c}^{-\frac{1}{2}})}\;da,
\end{equation}
and we find
\begin{equation}
B(\lambda=-1/8)=\frac{\pi}{2\Lambda_{c}G}\sqrt{\frac{3}{2}}\;
[\frac{17\sqrt{3}}{4}-\frac{7}{16}\mbox{ln}(4\sqrt{3}+7)]\approx
\frac{12}{\Lambda_{c}G} \propto C_{m}^2 .
\end{equation}
{}From eq.(3.7) we conclude that the probability of tunneling (3.1), is
inversely
related to the overall amount of nonrelativistic matter (dust): for a small
amount of nonrelativistic matter the tunneling probability approaches
unity, while it vanishes
asymptotically for a large amount of it. This is quite reasonable since
the quantum effects are thought to be connected with vacuum-like matter
domination (cosmological constant, strings, walls).

The plot of $B$ as a function of $\lambda$ in the full range
$\lambda\in\;]-1/8,\;0[,\;$ is shown in Fig.2. For small (negative) $\lambda,$
the relationship is essentially linear while $B$ increases steeply when
$\lambda$ approaches $-1/8.$

\section{Conclusion}
\setcounter{equation}{0}

In this paper we discussed the tunneling probability for a closed Friedmann
expanding universe, to change its evolution
completely and spontaneously and become an oscillating
universe, {\it without} any changes in the matter content. The tunneling
happens
merely when the closed universe reaches its maximum expansion size $a_{1}$
(Fig.1). It falls into the oscillating regime, beginning its further evolution
with a minimum value of the scale factor $a_{2}$ ($a_{2} > a_{1}$), and then
oscillates between $a_{2}$ and $a_{3}$. Eventually, after some period of
oscillations, when at the minimum $a_{2}$ again, the universe can tunnel
spontaneously {\it without} any changes in the matter content to the
recollapsing regime. It begins collapsing exactly with the same value of the
scale factor $a_{1}$ corresponding to the maximum size of the classical
Friedmann evolution. For the special solution of the Friedmann
equation with fixed
matter content of Section 2, we showed that the tunneling probability depends
on the (here necessarily negative) cosmological constant and it is bigger for
$\lambda$ closer to zero reaching unity for $\lambda = 0$. The latter case
corresponds to an unstable Einstein static solution together with the
asymptotic
solutions, see eq.(2.25) and Fig.1.

Also, we found that the tunneling probability is inversely related to the
amount of
nonrelativistic matter (dust). It means that for a universe filled with
more vacuum-type-matter the quantum effects are more probable and more
important. This is a quite reasonable result, since the quantum effects are
thought to be connected with vacuum-like matter domination (cosmological
constant, strings, walls).

Finally, one should mention the problem of arrow of time. Recently, Kiefer and
Zeh \cite{kieze,kie} claimed
that the arrow of time reverses just at the moment of
maximum expansion $a_{1}$ (Fig.1), in contradiction to the point of view of
Hawking and collaborators (e.g. \cite{haw,las}).
We do not discuss here the question,
whether there is a time asymmetry in our solutions, i.e., whether the expanding
solution before the tunneling and the collapsing solution after the
re-tunneling
are described by the different arrow of time directions. It seems to be matter
for further considerations.

\vskip 48pt
\hspace*{-6mm}{\bf Acknowledgements:}\\
A.L. Larsen is supported by the Danish Natural Science Research
Council under grant No. 11-1231-1SE
\vskip 48pt

\appendix
\section{The Euclidean Action}
\setcounter{equation}{0}

As was first shown by Coleman \cite{col},
the quantum mechanical tunneling probability
in the WKB-approximation,
can also be obtained by an instanton calculation. The probability is
\begin{equation}
p\sim e^{-S_{E}},
\end{equation}
where $S_{E}$ is the Euclidean action of the classical solution in the
classically forbidden region. In standard cases, the two expressions
(3.1) and (A.1) lead to the same result. In the case of tunneling Friedmann
universes, under consideration here, it is however instructive to perform the
computation in both ways. The reason is that when using (3.1), we do not
need an explicit expression for $S_{m}$ (under the assumption that
$\partial S_{m}/\partial a_{,t}=0$) and the computation proceeds as explained
in Section 3 with the result (3.7). When using now (A.1) instead, we
need an explicit expression
for the matter action $S_{m}.$ This is usually
not a problem when matter is represented
by scalar (or higher spin)
fields, but here the matter has been introduced only via
energy and pressure densities in the energy-momentum tensor (2.5). An
explicit expression for $S_{m}$ can still be obtained, but in a somewhat
tricky way by introducing an additional function $c(t)$ besides the scale
factor $a(t),$ see for instance Atkatz and
Pagels \cite{atp}. The action is written in the following way
\begin{eqnarray}
\tilde{S}_{g}\hspace*{-2mm}&=&\hspace*{-2mm}
\frac{3\pi}{4G}\int dt\;\frac{1}{c}
[-aa^2_{,t}+ac^2-\frac{\Lambda}{3}a^3c^2],
\nonumber\\
\tilde{S}_{m}\hspace*{-2mm}&=&\hspace*{-2mm}-\frac{3\pi}{4G}\int dt\;ca^3
[\frac{2}{3\sqrt{\Lambda_c}\;a^3}+\frac{3\sqrt{\Lambda_c}}{8a}].
\end{eqnarray}
$\tilde{S}_{g}$ reduces to eq.(3.4) for $c=1$ and $\tilde{S}_{m}$ was
chosen such that the Einstein equation (2.6), with the parameter values (2.10),
appears when taking variation with respect to $c$ (and taking $c=1$)
in the total action
$\tilde{S}=\tilde{S}_{g}+\tilde{S}_{m}.$ Taking variation with respect
to $a,$ (and taking $c=1$)
we get simply the time derivative of
the Einstein equation (2.6), with the parameter values (2.10).
For $c=1,$ and defining
$t\equiv it_{E},\;S_{E}\equiv i S,$ the Euclidean action is
\begin{equation}
S_{E}=\frac{3\pi}{4G}\int_{\mbox{period}} dt_{E}\;[a(\frac{da}{dt_{E}})^2+a-
\frac{\Lambda}{3}a^3-\frac{2}{3\sqrt{\Lambda_c}}
-\frac{3\sqrt{\Lambda_c}}{8}a^2].
\end{equation}
The Euclidean Einstein equation is
\begin{equation}
a^2(\frac{da}{dt_{E}})^2=\frac{-2a}{3\sqrt{\Lambda_{c}}}+a^2
-\frac{3\sqrt{\Lambda_{c}}}{8}a^3-\frac{\Lambda}{3}a^4.
\end{equation}
Now it is straightforward to convert the $t_{E}$-integral (A.3) into an
$a$-integral, and via (A.1) we recover the result, eqs.(3.1),(3.7),
for the tunneling
probability.

\section{Exact Elliptic Integral, eq.(3.7)}
\setcounter{equation}{0}
In this appendix, we give an explicit expression for the integral (3.7).
Defining
\begin{equation}
y^2=x(x-x_1)(x_2-x)(x_3-x),
\end{equation}
and using the reduction formulas
17.1.4 in Ref.\cite{as}, we get
\begin{eqnarray}
\int_{x_1}^{x_2}ydx\hspace*{-2mm}&=&\hspace*{-2mm}
-\frac{3}{32\lambda^2}
\int_{x_1}^{x_2}\frac{dx}{y}-\frac{1}{\lambda^3}(\lambda+\frac{81}{512})
\int_{x_1}^{x_2}\frac{dx}{xy}\nonumber\\
\hspace*{-2mm}&+&\hspace*{-2mm}
\frac{1}{\lambda^3}(\lambda^2+\frac{27\lambda}{32}+
\frac{729}{8192})\int_{x_1}^{x_2}\frac{xdx}{y}
\end{eqnarray}
These are standard elliptic integrals \cite{byrd}
\begin{eqnarray}
\int_{x_1}^{x_2}ydx\hspace*{-2mm}&=&\hspace*{-2mm}\frac{1}{\lambda^3
\sqrt{x_2(x_3-x_1)}}[\frac{-3\lambda}{16}+2x_3(\lambda^2+\frac{27\lambda}{32}+
\frac{729}{8192})-\frac{2}{x_3}(\lambda+\frac{81}{512})]K(m)\nonumber\\
\hspace*{-2mm}&+&\hspace*{-2mm}\frac{2(x_2-x_3)}{\lambda^3\sqrt{x_2(x_3-x_1)}}
[\lambda^2+\frac{27\lambda}{32}+\frac{729}{8192}]\Pi(\frac{x_1-x_2}{x_3-x_1}
\;|\;m)\nonumber\\
\hspace*{-2mm}&+&\hspace*{-2mm}\frac{2(x_2-x_3)}{\lambda^3 x_2 x_3
\sqrt{x_2(x_3-x_1)}}[\lambda+\frac{81}{512}]\Pi(\frac{x_3(x_1-x_2)}
{x_2(x_3-x_1)}\;|\;m),
\end{eqnarray}
in the notation of Abramowitz \cite{as}. The integral (3.7) is thus expressed
in terms of complete elliptic integrals of first and third kind.
The elliptic modulus is here given by
\begin{equation}
m\equiv k^2=\frac{x_3(x_2-x_1)}
{x_2(x_3-x_1)}.
\end{equation}

\newpage
\begin{centerline}
{\bf Figure Captions}
\end{centerline}
\vskip 72pt
\hspace*{-6mm}Fig.1. The potential $V(a),$ defined in eq.(2.24), for a
generic value of
$\lambda$ in the range $\;]-1/8,\;0[\;.$
The solution $a_{(-)}$ oscillates between
$a=0$ and $a=a_1$ (singular solution), while $a_{(+)}$ oscillates between
$a=a_2$ and $a=a_3$ (nonsingular solution).
\vskip 48pt
\hspace*{-6mm}Fig.2. The integral $B,$ defined in eq.(3.7), plotted here as a
function of $\lambda\in\;]-1/8,\;0[\;.$ The tunneling probability is
$p\sim\mbox{Exp}(-B).$

\end{document}